\documentclass[
preprintnumbers, 
showpacs, 
amsmath, 
showkeys, 
amssymb, 
tightenlines,
aps, 
superscriptaddress, 
11pt
]{revtex4-2}
\usepackage{lipsum, babel}
\usepackage{color}
\usepackage{hyperref}
\hypersetup{hidelinks}
\usepackage{bm}
\usepackage{amsfonts}
\usepackage{orcidlink}
\usepackage{mathrsfs}
\usepackage{booktabs}
\usepackage{graphicx}
\usepackage{amsmath}
\usepackage{float}
\usepackage{braket}
\usepackage{natbib}

     \hypersetup{
    colorlinks=true,
    linkcolor=red,
    citecolor=blue,
}

\begin{document}

\newcommand{\sgn}{\operatorname{sgn}}
\newcommand{\hhat}[1]{\hat {\hat{#1}}}
\newcommand{\pslash}[1]{#1\llap{\sl/}}
\newcommand{\kslash}[1]{\rlap{\sl/}#1}
\newcommand{\lab}[1]{}
\newcommand{\sto}[1]{\begin{center} \textit{#1} \end{center}}
\newcommand{\rf}[1]{{\color{blue}[\textit{#1}]}}
\newcommand{\eml}[1]{#1}
\newcommand{\el}[1]{\label{#1}}
\newcommand{\er}[1]{Eq.\eqref{#1}}
\newcommand{\df}[1]{\textbf{#1}}
\newcommand{\mdf}[1]{\pmb{#1}}
\newcommand{\ft}[1]{\footnote{#1}}
\newcommand{\n}[1]{$#1$}
\newcommand{\fals}[1]{$^\times$ #1}
\newcommand{\new}{{\color{red}$^{NEW}$ }}
\newcommand{\ci}[1]{}
\newcommand{\de}[1]{{\color{green}\underline{#1}}}
\newcommand{\ke}{\rangle}
\newcommand{\br}{\langle}
\newcommand{\lb}{\left(}
\newcommand{\rb}{\right)}
\newcommand{\lbk}{\left[}
\newcommand{\rbk}{\right]}
\newcommand{\blb}{\Big(}
\newcommand{\brb}{\Big)}
\newcommand{\nn}{\nonumber \\}
\newcommand{\p}{\partial}
\newcommand{\pd}[1]{\frac {\partial} {\partial #1}}
\newcommand{\cd}{\nabla}
\newcommand{\cc}{$>$}
\newcommand{\bqa}{\begin{eqnarray}}
\newcommand{\eqa}{\end{eqnarray}}
\newcommand{\bqe}{\begin{equation}}
\newcommand{\eqe}{\end{equation}}
\newcommand{\bay}[1]{\left(\begin{array}{#1}}
\newcommand{\eay}{\end{array}\right)}
\newcommand{\eg}{\textit{e.g.} }
\newcommand{\ie}{\textit{i.e.}, }
\newcommand{\iv}[1]{{#1}^{-1}}
\newcommand{\st}[1]{|#1\ke}
\newcommand{\at}[1]{{\Big|}_{#1}}
\newcommand{\zt}[1]{\texttt{#1}}
\newcommand{\m}{\mu}
\newcommand{\wei}[1]{\mathrm{d}#1}
\newcommand{\dao}[2]{\frac{\wei{#1}}{\wei{#2}}}
\newcommand{\pian}[2]{\frac{\partial #1}{\partial #2}}
\newcommand{\bian}[1]{\delta#1}
\newcommand{\bianfen}[2]{\frac{\bian{#1}}{\bian{#2}}}
\newcommand{\lrx}[1]{\left(#1\right)}
\newcommand{\lrz}[1]{\left[#1\right]}
\newcommand{\lrd}[1]{\left\{#1\right\}}
\newcommand{\non}{\nonumber\\}
\def\xa{{m}}
\def\xA{{m}}
\def\xb{{\beta}}
\def\xB{{\Beta}}
\def\xd{{\delta}}
\def\xD{{\Delta}}
\def\xe{{\epsilon}}
\def\xE{{\Epsilon}}
\def\xve{{\varepsilon}}
\def\xg{{\gamma}}
\def\xG{{\Gamma}}
\def\xk{{\kappa}}
\def\xK{{\Kappa}}
\def\xl{{\lambda}}
\def\xL{{\Lambda}}
\def\xo{{\omega}}
\def\xO{{\Omega}}
\def\xvp{{\varphi}}
\def\xs{{\sigma}}
\def\xS{{\Sigma}}
\def\xt{{\theta}}
\def\xvt{{\vartheta}}
\def\xT{{\Theta}}
\def \Tr {{\rm Tr}}
\def\CA{{\cal A}}
\def\CC{{\cal C}}
\def\CD{{\cal D}}
\def\CE{{\cal E}}
\def\CF{{\cal F}}
\def\CH{{\cal H}}
\def\CJ{{\cal J}}
\def\CK{{\cal K}}
\def\CL{{\cal L}}
\def\CM{{\cal M}}
\def\CN{{\cal N}}
\def\CO{{\cal O}}
\def\CP{{\cal P}}
\def\CQ{{\cal Q}}
\def\CR{{\cal R}}
\def\CS{{\cal S}}
\def\CT{{\cal T}}
\def\CV{{\cal V}}
\def\CW{{\cal W}}
\def\CY{{\cal Y}}
\def\BC{\mathbb{C}}
\def\BR{\mathbb{R}}
\def\BZ{\mathbb{Z}}
\def\sA{\mathscr{A}}
\def\sB{\mathscr{B}}
\def\sF{\mathscr{F}}
\def\sG{\mathscr{G}}
\def\sH{\mathscr{H}}
\def\sJ{\mathscr{J}}
\def\sL{\mathscr{L}}
\def\sM{\mathscr{M}}
\def\sN{\mathscr{N}}
\def\sO{\mathscr{O}}
\def\sP{\mathscr{P}}
\def\sR{\mathscr{R}}
\def\sQ{\mathscr{Q}}
\def\sS{\mathscr{S}}
\def\sX{\mathscr{X}}

	
	\title{Emergence of Photon Bose-Einstein Condensation from Down-Scattering in Cold Electron Media}
	    
	\author{Bingang Guo}
    \email{guobingang@impcas.ac.cn}
    \affiliation{Institute of Modern Physics, Chinese Academy of Sciences, Lanzhou 730000, China}
    \affiliation{School of Nuclear Science and Technology, University of Chinese Academy of Sciences, Beijing 100049, China}
\author{Wei Kou\orcidlink{0000-0002-4152-2150}}
\email{kouwei@impcas.ac.cn}
\affiliation{Institute of Modern Physics, Chinese Academy of Sciences, Lanzhou 730000, China}
\affiliation{School of Nuclear Science and Technology, University of Chinese Academy of Sciences, Beijing 100049, China}
\affiliation{Southern Center for Nuclear Science Theory (SCNT), Institute of Modern Physics, Chinese Academy of Sciences, Huizhou 516000, Guangdong Province, China}
\author{Xurong Chen}
\email{xchen@impcas.ac.cn (corresponding\,author)}
\affiliation{Institute of Modern Physics, Chinese Academy of Sciences, Lanzhou 730000, China}
\affiliation{School of Nuclear Science and Technology, University of Chinese Academy of Sciences, Beijing 100049, China}
\affiliation{Southern Center for Nuclear Science Theory (SCNT), Institute of Modern Physics, Chinese Academy of Sciences, Huizhou 516000, Guangdong Province, China}
\begin{abstract}
In this study, we examine the emergence of photon Bose-Einstein condensation (BEC) resulting from the interaction of high-energy photons with a cold electron gas, modeled via a modified Kompaneets equation. Beginning with an initial black-body photon spectrum, we perform numerical simulations to track the evolution of the photon distribution under the influence of inverse Compton scattering, wherein photons dissipate energy through collisions with cold electrons. Our results demonstrate a pronounced enhancement of photon number density at the low-energy tail, indicative of a BEC-like phase transition. This phenomenon is further corroborated by an analysis of the entropy evolution during the cooling process, revealing that the condensate configuration corresponds to the entropy maximum, in accordance with thermodynamic principles. These findings establish a comprehensive theoretical framework for photon BEC formation in cold electron environments and underscore the significance of entropy maximization as a driving mechanism for condensation.
\end{abstract}	
	\maketitle
\section{Introduction}
	Elementary particles are classified into fermions and bosons according to statistics, with bosons following Bose-Einstein statistics. In the 1920s, predictions were made by Satyandra Nath Bose  \cite{1924ZPhy...26..178B} and Albert Einstein based on Bose's statistical mechanics study of photons. According to this statistical law, it is predicted that when the temperature approaches absolute zero, bosons will aggregate within the same energy range, forming a gaseous, superfluid state of matter known as the Bose-Einstein condensate (BEC). Previous experiments often utilized atoms as the smallest unit to investigate this phenomenon. Photons, being the most common boson, are expected to be capable of forming a BEC under specific conditions. Due to increased interference in ground-based laboratories, there have been no reliable experiments to confirm this process. However, in the universe, the presence of cold electrons or plasma in the early universe suggests that photons at higher temperatures will interact with electrons and other substances, leading to their cooling. Consequently, the universe is considered a promising environment for observing BEC phenomena.
	
	In the early universe, it is known to have abundant particle fluids such as electron gas and plasma. In particular, the cosmic microwave background (CMB) radiation is present. Therefore, CMB photons will inevitably undergo strong Compton scattering with electrons or plasma. The interaction of CMB radiation with electrons through Compton scattering can be described by the classical Kompaneets equation  \cite{Kompaneets:1957kya},
\begin{equation} 
	\left(\frac{\partial n}{\partial t}\right)_c=\frac{kT_e}{m_ec^2}N_e\sigma_{\mathrm{T}}c\frac{1}{x^2}\frac{\partial  }{\partial x}\left\{x^4\left[\frac{\partial n}{\partial x}+n(n+1)\right]\right\},
	\label{eq:okeq}
\end{equation}
where $x\equiv h\nu/kT_e$ represents the dimensionless energy of a photon with the electron temperature $kT_e$; $N_e$ denotes the number density of the scattering electrons; $n(x,t)$ is the energy distribution function of the CMB photons, and $\sigma_{\mathrm{T}}$ is the Thomson cross-section. The Kompaneets equation is commonly used to describe the up-Comptonization scattering of low-energy photons with $h\nu\ll kT_e$. However, it is not suitable for describing the down-Comptonization process of high-energy or high-temperature photons, which is an important radiation process in hard $\gamma$-ray astronomy. As mentioned, the classical equation can only depict a scenario where the initial electron temperature is higher than the photon temperature, resulting in the photon being in a heated state, causing its spectral peak to shift towards higher energies and gradually broaden. Conversely, if the electron temperature is initially lower than the photon temperature, the electron cools the photon to a lower energy level, leading to the generation of a pulse near the origin after a certain time, known as a 
BEC  \cite{Zeldovich:1969kya}.

	\begin{widetext}
	Although the case of $T_e < T_\gamma$ can be directly considered in the classical equation, there is a technical logical error. Therefore, based on the classical equation, an modified Kompaneets equation has been derived \cite{chen2021modified} following the method of Ross and McCray \cite{ross1978comptonization}, which not only satisfies photon number conservation but is also applicable to the situation where $T_e < T_\gamma$. The modified Kompaneets equation is written as

\begin{equation} 
	\left(\frac{\partial n}{\partial t}\right)_c=\frac{kT_e}{m_ec^2}N_e\sigma_{\mathrm{T}}c\frac{1}{x^2}\frac{\partial }{\partial x}\bigg\{x^4\left(1+\frac{14}{5}\frac{kT_e}{m_ec^2}x\right)
	\times\left[\frac{\partial n}{\partial x}+n(n+1)\right]\bigg\}.
	\label{eq:mkeq}
\end{equation}	
\end{widetext}
Compared to the classical Kompaneets equation (\ref{eq:okeq}), an additional correction term dependent on $x$ is present on the right-hand side of the Eq. (\ref{eq:mkeq}), which can be neglected when photon energy is low, but becomes significant in the case of high-energy photons.
It is straightforward to demonstrate that both the classical and the modified Kompaneets equations share the same stationary solution $n_\mu(x)=(e^{x-\mu}-1)^{-1}$. Many researchers have worked on obtaining the exact solution of the classical Kompaneets equation. Nagirner \cite{nagirner1997exact} employed the method of Green's function to analytically solve the linear Kompaneets equation and presented a numerical difference decomposition of the corresponding equation. Bluman and collaborators \cite{2014JEnMa..84...87B} analyzed the properties of the nonlinear Kompaneets equation based on group theory and provided analytic solutions. In 2016, Pego and colleagues \cite{doi:10.1137/15M1054730} derived a nonlinear hyperbolic partial differential equation while investigating the classical Kompaneets equation, ultimately demonstrating the appearance of a BEC after a certain evolution period. For the numerical solution of this equation, Chluba and Sunyaev \cite{10.1111/j.1365-2966.2011.19786.x,Chluba:2013vsa} developed a program package called C\begin{scriptsize}OSMO\end{scriptsize}T\begin{scriptsize}HERM\end{scriptsize}  based on the Monte Carlo method and Green's function to directly solve the Kompaneets equation when studying the evolution of the early CMB, yielding results in good agreement with experimental data. BEC finds wide applications in condensed matter physics and astrophysics. In this study, the characteristic line method \cite{Zeldovich:1969kya} is employed to obtain the analytic solution after simplifying the modified Kompaneets equation. Comparison with the numerical solution reveals that similar BEC results can be achieved.

	In the late 1960s, the space X-ray telescope discovered that galaxy clusters were diffuse X-ray sources, which was interpreted as the bremsstrahlung of the vilified high-temperature ionized gas within the cluster. The Soviet astrophysicists Sunyaev and Zel'dovich \cite{Zeldovich:1969ff,Sunyaev:1970er} immediately proposed that free electrons in ionized gas would scatter the passing CMB photons, causing distortion of the CMB blackbody energy spectrum, reducing the energy intensity at the low frequency end and increasing the high frequency end. This is the famous Sunyaev-Zel'dovich (SZ) effect. The SZ effect is a phenomenon that can be observed in cosmological experiments. The CMB spectrum will be distorted when the scattering time of photons and electrons is very small. 
	When the scattering time is long enough for the photons to agglomerate, the spectrum evolves into BEC. 
	Therefore, the SZ effect is a starting point for observing the BEC of photons in the universe. In the study of nucleon structure, protons are composed of quarks and gluons, which belong to gauge bosons and satisfy Bose-Einstein statistics. If gluon condensation can be explained by BEC, it will be a bright prospect in future. In this article, we followed the work of Y. B. Zel'dovich and E. V. Levich \cite{Zeldovich:1969kya} to make a simple approximation to find the analytical solution of the modified equation, and we can verify that the numerical solution and the analytical solution are self-consistent within a certain range, and both can explain physical phenomena. Finally, we give the results of SZ effect from modified version and discuss it.

\section{Modified Kompaneets Equation and Analytical Interpretation}

	The modified Kompaneets equation was derived by using $\Delta p$ rather than $\Delta \nu$ in the Taylor expansion \cite{Zhang:2015kca}. Therefore, compared to the original Kompaneets equation, the modified one is applicable not only to the case where $h\nu\ll kT_e$, but also to the case where $h\nu\simeq kT_e$, and in particular to the case where $h\nu\gg kT_e$.
	First, we make physical approximation of the equation based on the work of Zel'dovich and Levich \cite{Zeldovich:1969kya}, and we get reliable analytical results. Secondly, we directly solve the equation numerically, which can finally explain the analytical results and the real physical process.

	With the modified equation:
\begin{equation} 
\pian{n}{y}=\frac{1}{x^2}\frac{\partial}{\partial x}\left\{x^4(1+Ax)\left(\frac{\partial n}{\partial x}+n^2+n\right)\right\},
\label{sim-eke}
\end{equation}
	where $y=kT_eN_e\sigma_{\text{T}}t/(m_ec)$, $A=14k_BT_e/(5m_ec^2)$. In the case $T_e<T_\gamma$, the Doppler boosting of low-energy photons to higher frequency is weaker than the down scattering of high-energy photons by recoil and induced recoil. Therefore, for $y\ll 1$ 
	by approximating $\partial n/\partial x + n\ll n^2$  \cite{Zeldovich:1969ff}, Eq.~(\ref{sim-eke}) can be rewritten as:
\begin{equation} 
	\frac{\partial f}{\partial y}-2f(1+Ax)\frac{\partial f}{\partial x}=Af^2.
	\label{app-eke}
\end{equation}
	where $f=x^2n$. Eq.~(\ref{app-eke}) is a quasi-linear partial differential equation, which can be solved by method of characteristics.The corresponding Lagrange-Charpit equations read:
\begin{equation}
	\frac{\mathrm{d}y}{1}=\frac{\mathrm{d}x}{-2f(1+A x)}=\frac{\mathrm{d}f}{A f^2}.
	\label{lce}
\end{equation}
According to the first Lagrange-Charpit equation, slope of the characteristic curves is obtained:
\begin{equation}
    \dao{x}{y}=-2f(1+Ax).
    \label{slope}
\end{equation}
Eq.~(\ref{slope}) indicates the speed of high-energy photons moving towards low frequency. Integrating the rest two Lagrange-Charpit equations, one botain:
\begin{equation}
            -f\sqrt{1+A x}=\Phi,~A y+\frac{1}{f}=\Delta,
    \end{equation}
    where $\Phi$ and $\Delta$ are integration constants. A family of one-parameter characteristic curves is defined by assuming $\Phi$ as a function of $\Delta$, $\Phi=g(\Delta)$. A general solution for Eq.~(\ref{app-eke}) can be written implicitly as:
    \begin{equation}
        \begin{aligned}
            f\sqrt{1+A x}=g\lrx{A y+\frac{1}{f}}.
        \end{aligned}
        \label{gensol}
    \end{equation}

\begin{figure}[htbp]
    \centering
    \includegraphics[scale=0.48]{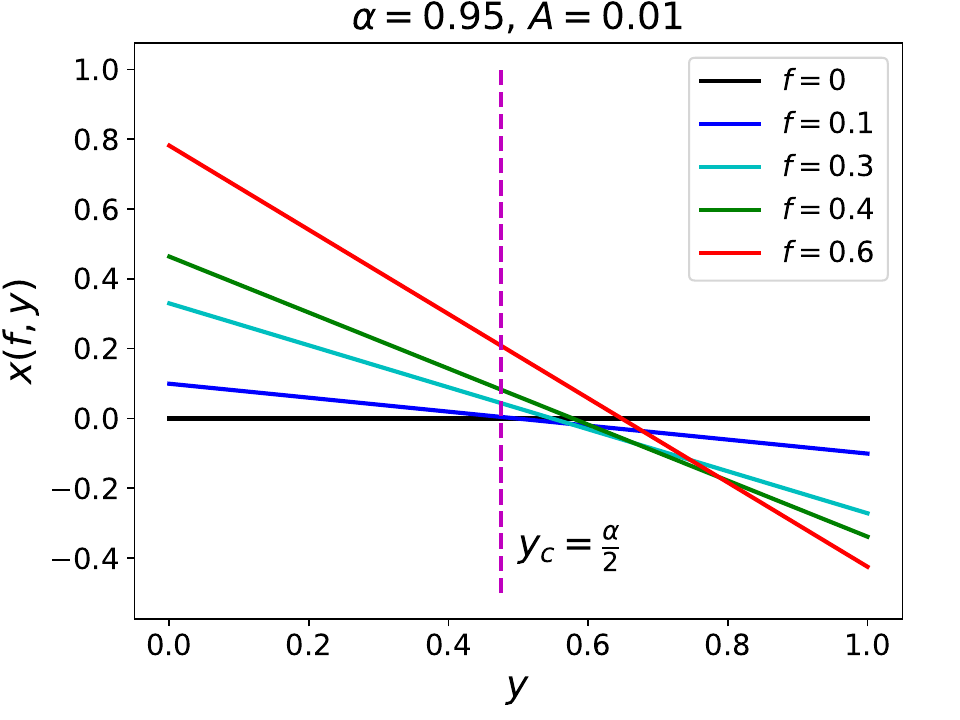}
    \caption{(color online) Evolution of positions of photons in momentum space.}
    \label{fig-app-sol-x}
    \end{figure}

	For an initial blackbody distribution at temperature $T_\gamma$, at low frequency, one obtain the approximate initial condition 
	\begin{equation}
	    \lim_{x\to 0+}f(x,0)
	    =\lim_{x\to 0+}\frac{x^2}{e^{\alpha x}-1}\simeq\frac{2x}{2\alpha + \alpha^2 x},
	\label{inicon}
\end{equation}
where $\alpha\equiv T_e / T_\gamma < 1$.	Taking Eq.~(\ref{inicon}) into Eq.~(\ref{gensol}), one obtain
	\begin{equation}
        y f + \frac{x}{2} = \frac{\alpha f}{2(A y f+1) - \alpha^2f},
        \label{imp-sol}
    \end{equation}
    from Eq.~(\ref{imp-sol}), one gets
    \begin{equation}
        x(f,y)=-\frac{2 \left(-2 A f^2 y^2+\alpha ^2 f^2 y+\alpha  f-2 f y\right)}{\alpha^2 f-2 A f y-2},
        \label{eq-app-sol-x}
    \end{equation}
    which describes the evolution of positions of photons in the momentum space. Moreover, one can see that there are no photons in the unphysical region $x<0$ until a critical time $y_c=\alpha/2$ (See Fig. \ref{fig-app-sol-x}).
    
    \begin{widetext}
    Consequently, an approximate solution can be given:
    \begin{equation}
        f=\frac{4 y - 2 \alpha - \alpha^2 x + 2 A y x \pm \sqrt{16 (\alpha^2 y - 2 A y^2) x + (4 y - 2 \alpha - \alpha^2 x + 2 A y x)^2}}{4 (\alpha^2 y - 2 A y^2)},~y < \frac{\alpha^2}{2 A}.
        \label{app-sol}
    \end{equation}
\end{widetext}

    \begin{figure}[htbp]
        \centering
        \includegraphics[scale=0.48]{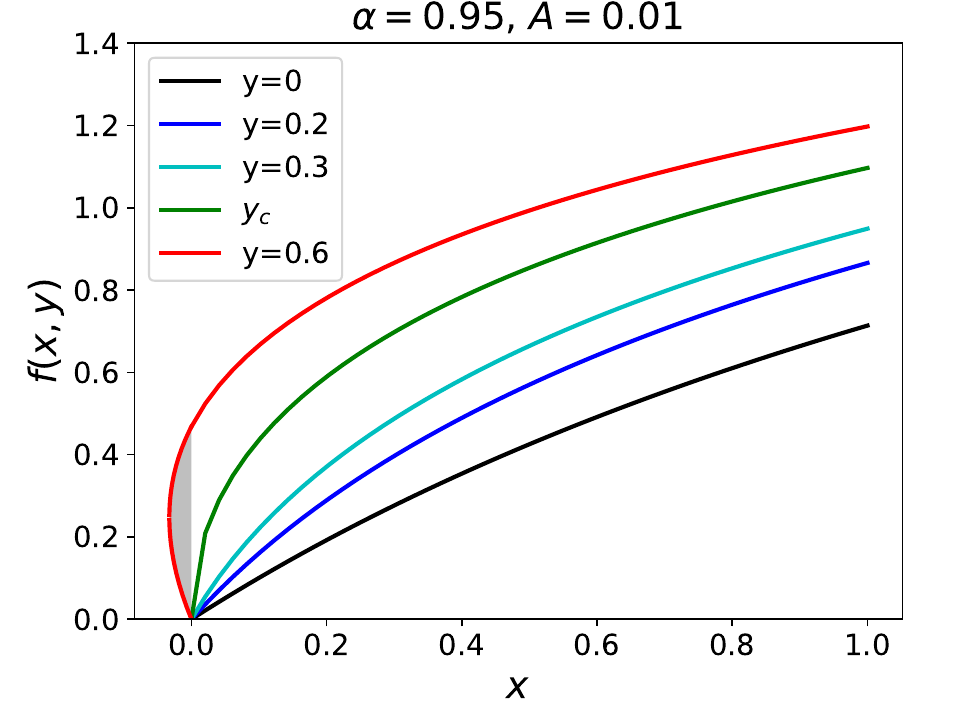}
        \caption{(color online) Approximate solution to the modified Kompaneets equation with $A=0.01$ and $\alpha=0.95$. Shaded area denotes photons in the unphysical region.}
        \label{fig-app-sol}
    \end{figure}
    
    Fig.~\ref{fig-app-sol} illustrates the approximate photon distribution function at different time $y$. According to the approximate solution, the initial distribution function evolves with photons move towards lower frequency. The solution (\ref{app-sol}) formally indicates that photons may move into the negative-energy region (as the shaded area shown in Fig.~\ref{fig-app-sol}). However, the transition to negative energy is impossible. Therefore, photons can only accumulate in the zero-energy state, which forms the Bose-Einstein Condensation.
    
	\section{Entropy and Bose-Einstein Condensation}
	In the above section, we discuss the photon spectra redistribution approximately by applying the method of characteristics to get approximate solution to the modified Kompaneets equation. On the other hand, the photon gas and thermal plasma form a larger thermodynamic system. As the photon spectra evolves, the full system changes as well, which makes it possible to see the effects of the Compton scattering from a different point of view. Here, we investigate how does the thermodynamics entropy of the full system evolve during the interaction.
	
	Dimensionless thermodynamic entropy of a Bose-Fermi interaction system is defined as functional of the distribution of photons:
	\begin{equation}
    \begin{aligned}
        S[n]
        &=\int\mathrm{d}x s[n]x^2\\
        s[n]
        &=(1+n)\ln (1+n)-n\ln n-xn.
    \end{aligned}
    \label{entropy}
\end{equation}
where $s[n]$ denotes the density of the system entropy. And the derivatives:
\begin{equation}
	\begin{aligned}
		        \partial_n s &= \ln \left(1+\frac{1}{n}\right)-x,\\
		\partial^2_{n} s &= - \frac{1}{n + n^2},\\
		\partial_x\left(\partial_n s\right) &= - \frac{\partial_x n + n + n^2}{n + n^2}.
	\end{aligned}
\end{equation}
So $\left.\partial_n s\right|_{n_0}=0$, for any non-negative $n$ and $x$, one have 
$s[n]\leq s[n_0]=\pi^4/45\approx 2.16$. 

Hence, the evolution of entropy $S[n(x,t)]$ under the Kompaneets equations is:
\begin{equation}
    \begin{aligned}
    \frac{\mathrm{d} S}{\mathrm{d} y}
    =&\int\mathrm{d}x\frac{\mathrm{d}s}{\mathrm{d} y}x^2,\\
    =&\int\mathrm{d}x\left(\partial_n s\right)\left(\partial_y n\right) x^2,\\
    =&\int\mathrm{d}x\left(\partial_n s\right)\partial_x\left[W(x)\left(\partial_x n + n + n^2\right)\right],\\
    =&-\int\mathrm{d}x W(x)\left(\partial_x n + n + n^2\right)\partial_x\left(\partial_n s\right),\\
    =&\int\mathrm{d}x \left(\partial_x n + n + n^2\right)^2 \frac{W(x)}{n + n^2}\geq 0,
    \end{aligned}
    \label{evo-ent}
\end{equation}
where $W(x)=x^4$ for the classical Kompaneets equation and $W(x)=x^4(1+Ax)$ for the modified equation. Eq.~(\ref{evo-ent}) indicates that for the Kompaneets equations, the entropy of the radiation-plasma system increases with time increasing. 
Therefore, according to the principal of maximum entropy, the photons tend to approach the Planck distribution $n_0(x)$.

According to Ref.~ \cite{rusel:1986},
\begin{equation}
    S[n]=S\left[n_0 + \left| N - N_0\right| \delta(x)x^{-2}\right],
    \label{theorem}
\end{equation}
\begin{figure}[htbp]
\centering
\includegraphics[scale=0.48]{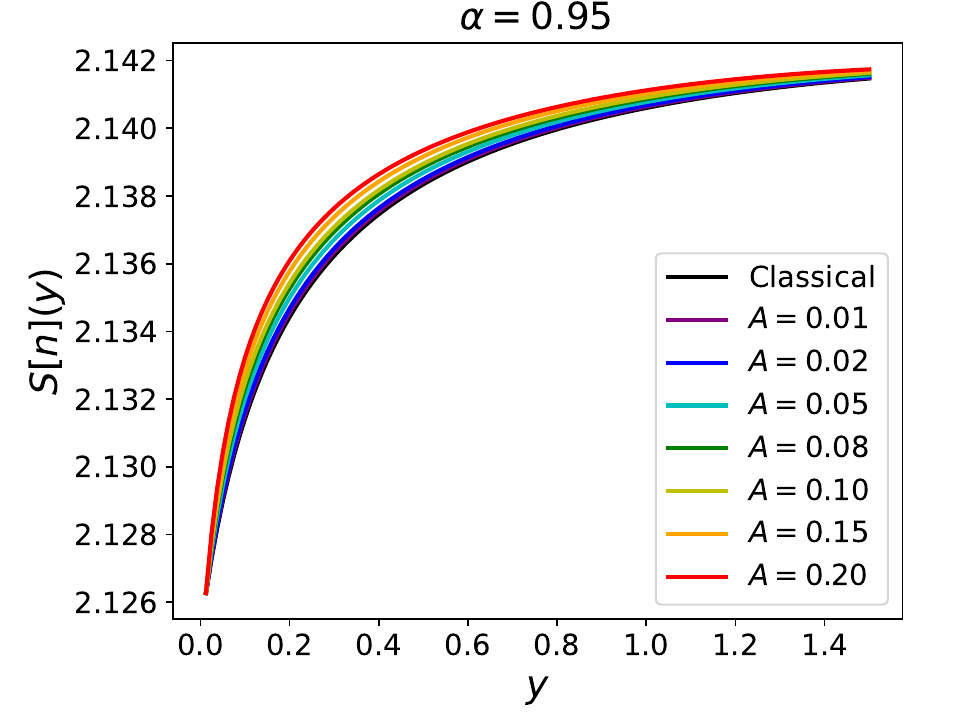}
\caption{(color online) Time evolution of the system entropy $S[n](y)$}
\label{fig-evo-ent}
\end{figure}
where $\delta(x)$ is the Dirac delta function. Eq.~(\ref{theorem}) indicates that it is possible for a system containing more photons than $N_0$ to achieve the maximum of entropy.

	In the previous section, an approximate solution of the modified Kompaneets equation Eq.~(\ref{sim-eke}) is presented with the key assumption that the induced recoil term dominant when $T_e < T_\gamma$. 
	Since both the Kompaneets equation and the modified equation are proposed to describe redistribution of photons in momentum space during the Compton scattering, as clarified previously, the number of photons must be conserved.
	\begin{equation}
	    0=\frac{\text{d}}{\text{d} y}N(y)\equiv\frac{\text{d}}{\text{d} y}\int\wei{x} f(x,y).
	\end{equation}
	For the stationary solution $n_\mu(x)=(e^{x-\mu} - 1)^{-1}$, the total photon number is
	\begin{equation}
	    \begin{aligned}
	      N_\mu
	      &\equiv\int_0^\infty\mathrm{d}x\frac{x^2}{e^{x-\mu} - 1}<\int_0^\infty\mathrm{d}x\frac{x^2}{e^x - 1}=N_0,  
	    \end{aligned}
	\end{equation}
	hence, if one set the Planck distribution (\ref{inicon}) to be the initial condition, the total photon number is
	\begin{equation}
		\begin{aligned}
			N
			&=\int_0^\infty\mathrm{d}x\frac{x^2}{e^{\alpha x} - 1} \geq N_0.
		\end{aligned}
	\end{equation}
	It means that there are more photons in the initial state than a stationary solution $N_\mu$.
	Consequently, the evolution for an initial Planck distribution will results in photons accumulating in the zero-energy state.
	
\begin{figure}[htbp]
	\centering
	\includegraphics[width=0.48\textwidth]{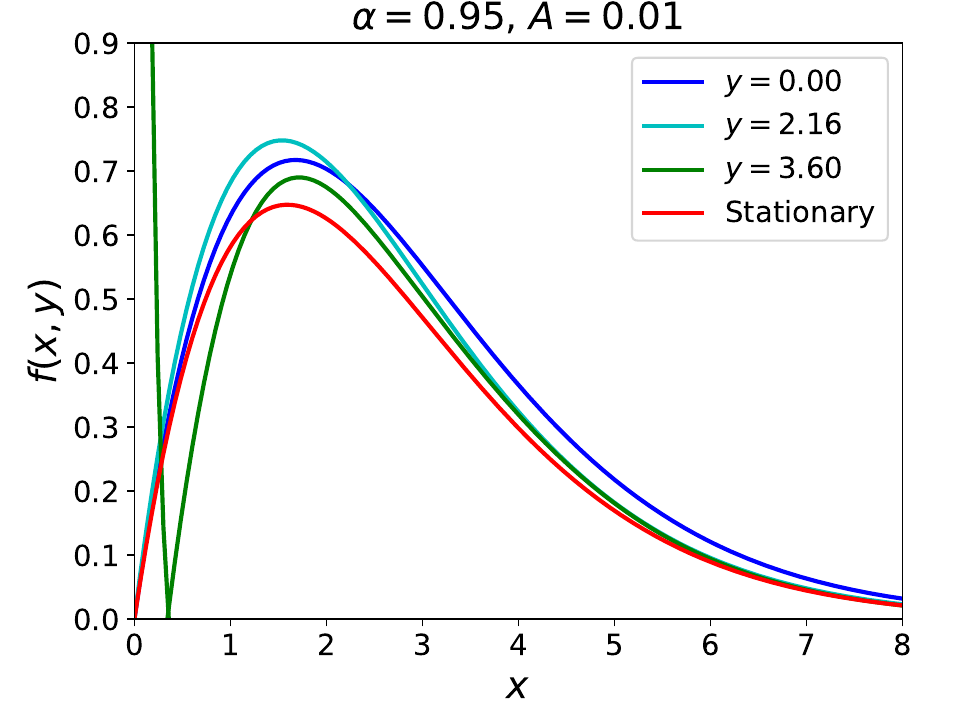}
	\caption{(color online) Numerical results for the modified Kompaneets equation over time, with $A=0.01$ and $\alpha=0.95$. The red dotted line denotes a state that the BEC of CMB photons has been developed, the residual photons tend to achieve the Planck distribution at temperature $T_e$.}
	\label{fig-num-sol}
	\end{figure}

	It is now worthwhile to compare the analytical results obtained from the EKE against those from full numerical simulations. We map the numerical solution of Eq.~(\ref{sim-eke}) onto the photon distribution function, $f = x^2 n$ and examine its evolution with respect to the dimensionless time variable $y$. The results for the parameter set $A = 0.01$ and $\alpha = 0.95$ are shown in Fig. \ref{fig-num-sol}. Notably, at certain values of $y$, the low-$x$ region—corresponding to low‑energy photons—exhibits singular‑like behavior. This behavior is indicative of photon condensation under the constraint of photon number conservation. Moreover, the photon–electron scattering dynamics that govern the evolution lead to a softening of the photon spectrum, resulting in a distribution that resembles BEC.
	
\section{Discussion and Conclusion}
Based on the preceding calculations and discussions, it is evident that within the high-photon-energy scattering regime, our generalized Kompaneets equation achieves a superior description-compared to the original Kompaneets formulation-of the photon gas’s momentum-space redistribution following Compton scattering with a thermal plasma. The new approximation retains a first-order momentum correction term, which becomes negligible in the low-photon-energy limit, thereby ensuring compatibility with the standard equation. However, at relatively high photon energies, this correction term has a non-negligible impact on the evolution of the photon spectrum. While the presence of the correction does not alter the stationary momentum-space distribution (i.e., the Bose-Einstein distribution) ultimately reached by the photon gas, it significantly influences the process by which this steady-state distribution is established.

In this work, we employ the method of characteristics to obtain an approximate solution for the behavior of the modified Kompaneets equation in the limit $x\to 0$
. By approximating the second-order equation as a first-order one and subsequently solving its characteristic equations, we can effectively capture the evolution of the distribution function. As shown in Figs. \ref{fig-app-sol-x} and \ref{fig-app-sol}, as the dimensionless time variable $y$ increases, the envelope of the distribution function shifts toward negative $x$. Since $x$ cannot be negative and the total photon number must be conserved, photons that would nominally occupy the $x<0$ region instead accumulate at $x=0$, giving rise to the BEC.

The conclusions drawn from the approximate analytic solution are likewise confirmed through numerical integration of the modified Kompaneets equation. As shown in Fig.~\ref{fig-num-sol}, over the course of time, the photon distribution evolves from its initial Maxwellian form at temperature $T_\gamma$ toward a stationary Bose-Einstein distribution characterized by the electron temperature $T_e$. During this process, the most probable photon energy steadily decreases, and a non-negligible photon population builds up at zero energy, signaling the formation of the BEC.

On the other hand, under the constraint of photon number conservation, the full system of photon gas interacting with the plasma undergoes a spontaneous increase in thermodynamic entropy. To preserve this entropy-increasing process, a subset of photons naturally accumulates at the zero-energy position, $x=0$. In other words, within the regime described by the modified Kompaneets equation, photon Bose–Einstein condensation and the second law of thermodynamics are mutually consistent.

In summary, in the quasi-classical Compton scattering regime, we have pursued solutions to the modified Kompaneets equation from two complementary perspectives: analytic approximations and numerical solutions. The formulation of the modified Kompaneets equation does not incorporate explicitly quantum effects beyond particle discreteness; it treats the photon distribution function’s evolution with classical scattering methods, yet the solution naturally encompasses BEC behavior. This suggests that interactions between cosmic rays and interstellar dust in space may tend to form photon BEC. On the other hand, in realistic interaction scenarios, number‑nonconserving processes such as double Compton scattering and bremsstrahlung can disrupt this tendency, potentially precluding observational signatures of cosmic‑ray–induced BEC. Nonetheless, we believe this kinematically induced BEC phenomenon merits further investigation.

\begin{acknowledgements}
	 This work is supported by the National Key R$\&$D Program of China under the Grant NO. 2024YFE0109800 and 2024YFE0109802.
\end{acknowledgements}
\bibliographystyle{unsrt}
\bibliography{ref}

\begin{thebibliography}{10}

\bibitem{1924ZPhy...26..178B}
{S. N. Bose}.
\newblock {\em {Zeitschrift fur Physik}}, {26}({1}):{178--181}, 1924.

\bibitem{Kompaneets:1957kya}
{\relax A}.~S. Kompaneets.
\newblock {\em Sov. Phys. JETP}, 4:730--737, 1957.
\newblock [J. Exptl. Theoret. Phys. (U. S. S. R)31, 876-885 (November, 1956)].

\bibitem{Zeldovich:1969kya}
{\relax Ya}.~B. Zeldovich and E.~V. Levich.
\newblock {\em Sov. Phys. JETP}, 28:1287--1290, 1969.
\newblock [Zh. Eksp. Teor. Fiz.55,2423(1969)].

\bibitem{chen2021modified}
Xurong Chen, Xu~Zhang, Haixiang Gao, Chengdong Han, and Dangbo Liu.
\newblock Modified form of the kompaneets equation-possible application to the
  distortion of high-temperature blackbody and cmb spectra.
\newblock {\em Astronomy \& Astrophysics}, 650:A74, 2021.

\bibitem{ross1978comptonization}
Randy~R Ross, Robert Weaver, and Rrchard McCray.
\newblock {\em The Astrophysical Journal}, 219:292--299, 1978.

\bibitem{nagirner1997exact}
DI~Nagirner, VM~Loskutov, and SI~Grachev.
\newblock {\em Astrophysics}, 40(3):227--236, 1997.

\bibitem{2014JEnMa..84...87B}
George~W. Bluman, Shou-fu Tian, and Zhengzheng Yang.
\newblock {\em Journal of Engineering Mathematics}, 84(1):87--97, 2014.

\bibitem{doi:10.1137/15M1054730}
Joshua Ballew, Gautam Iyer, and Robert~L. Pego.
\newblock {\em SIAM Journal on Mathematical Analysis}, 48(6):3840--3859, 2016.

\bibitem{10.1111/j.1365-2966.2011.19786.x}
J.~Chluba and R.~A. Sunyaev.
\newblock {\em Monthly Notices of the Royal Astronomical Society},
  419(2):1294--1314, 2011.

\bibitem{Chluba:2013vsa}
Jens Chluba.
\newblock {\em Mon. Not. Roy. Astron. Soc.}, 434:352, 2013.

\bibitem{Zeldovich:1969ff}
{\relax Ya}.~B. Zeldovich and R.~A. Sunyaev.
\newblock {\em Astrophys. Space Sci.}, 4:301--316, 1969.

\bibitem{Sunyaev:1970er}
R.~A. Sunyaev and {\relax Ya}.~B. Zeldovich.
\newblock {\em Astrophys. Space Sci.}, 7:20--30, 1970.

\bibitem{Zhang:2015kca}
Xu~Zhang and Xurong Chen.
\newblock arxiv:1509.00140.
\newblock 2015.

\bibitem{rusel:1986}
Russel E. Caflisch; C.~David Levermore.
\newblock {Equilibrium for radiation in a homogeneous plasma}.
\newblock {\em Phys. Fluids}, 29:78, 1986.

\end{thebibliography}
\end{document}